\documentclass[conference,letterpaper]{IEEEtran}
\ifCLASSINFOpdf
 \else
\fi

\usepackage{layout}
\usepackage{wrapfig}
\usepackage{comment}
\usepackage{balance}  
\usepackage{cite}
\usepackage{graphicx}
\usepackage{float}
\usepackage{algorithm}
\usepackage{algpseudocode}
\usepackage{amsmath} 
\usepackage[utf8]{inputenc}
\usepackage[final]{pdfpages} 
\usepackage{pdfpages}
\usepackage{lipsum}
\usepackage{textcase}
\usepackage{url}
\usepackage{amsmath,esint}
\usepackage{epstopdf}
\usepackage{array} 
\usepackage[hidelinks]{hyperref}
\usepackage{multirow}
\usepackage{booktabs}
\usepackage{esvect}
\usepackage{soul}
\usepackage[normalem]{ulem}
\usepackage{amsbsy}
\usepackage{amssymb}
\usepackage{subcaption}

\usepackage[
  letterpaper,
  left=0.625in,
  right=0.6245in,
  top=0.74in,
  bottom=1.04in,
  columnsep=0.245in
]{geometry}

\usepackage{siunitx}  

\usepackage{booktabs} 
\usepackage{adjustbox} 

\newcolumntype{P}[1]{>{\centering\arraybackslash}p{#1}}
\newcolumntype{M}[1]{>{\centering\arraybackslash}m{#1}}

\begin{document}
%

\title{Propagation Channel Modeling for LEO Satellite Missions Using Ray-Tracing Simulations }

\author{\IEEEauthorblockN{Wahab Khawaja\IEEEauthorrefmark{1},~ 
 Ismail Guvenc\IEEEauthorrefmark{2}, and Rune Hylsberg Jacobsen\IEEEauthorrefmark{1}
}

\IEEEauthorblockA{\IEEEauthorrefmark{1}Dept. Electrical and Computer Engineering, Aarhus University, 8000 Aarhus, Denmark}

\IEEEauthorblockA{\IEEEauthorrefmark{2} Dept. Electrical and Computer Engineering, North Carolina State University, Raleigh, NC 27606, USA}

Email: wahabgulzar@ece.au.dk, iguvenc@ncsu.edu, rhj@ece.au.dk
}

\maketitle

\begin{abstract}
This work presents a high-resolution, ray-tracing-based channel modeling for Low Earth Orbit~(LEO) satellite-to-ground links in a suburban environment at X-band. Using simulations conducted in Wireless InSite, we develop a parametric channel model that characterizes both large- and small-scale fading effects across different satellite elevation angles. Large-scale fading incorporates attenuation due to terrain-induced shadowing and dynamic environmental factors such as weather conditions, and is compared with $3$GPP NTN channel model. Additionally, we quantify link degradation resulting from ground station~(GS) antenna misalignment, considering both fixed single-element and electronically steerable phased-array antennas. Small-scale fading is modeled by fitting a shadowed and non-shadowed Rician distribution to the fading statistics at various satellite elevations. To the best of our knowledge, this is the first study to propose a comprehensive elevation-aware channel model for satellite-to-ground propagation at X-band, integrating ray-traced environmental dynamics, elevation-dependent fading, and phased-array beam misalignment effects. 
\end{abstract}


\IEEEpeerreviewmaketitle

\section{Introduction}

Satellite communication has gained significant importance for current and future communication networks~\cite{sat_comm} and is expected to become part of $5$G and beyond networks~\cite{sat_5G, sat_5G_2}. Other expanding areas of the satellite communication market include navigation and remote sensing, Internet access to remote areas, defense, space exploration, and the Internet of things~\cite{sat_app}. Small satellites are playing an increasingly important role in advancing satellite communications and offer concise and cost-effective solutions that drive innovation in both research and commercial applications. These satellites have enabled applications such as Earth observation~\cite{disco2} and climate monitoring by small and medium-sized institutes worldwide~\cite{micro_cubesat, survey_cubesat}. 

\begin{figure}[ht!]
    \centering
    \begin{subfigure}[b]{0.48\textwidth} 
        \centering
        \includegraphics[width=\textwidth]{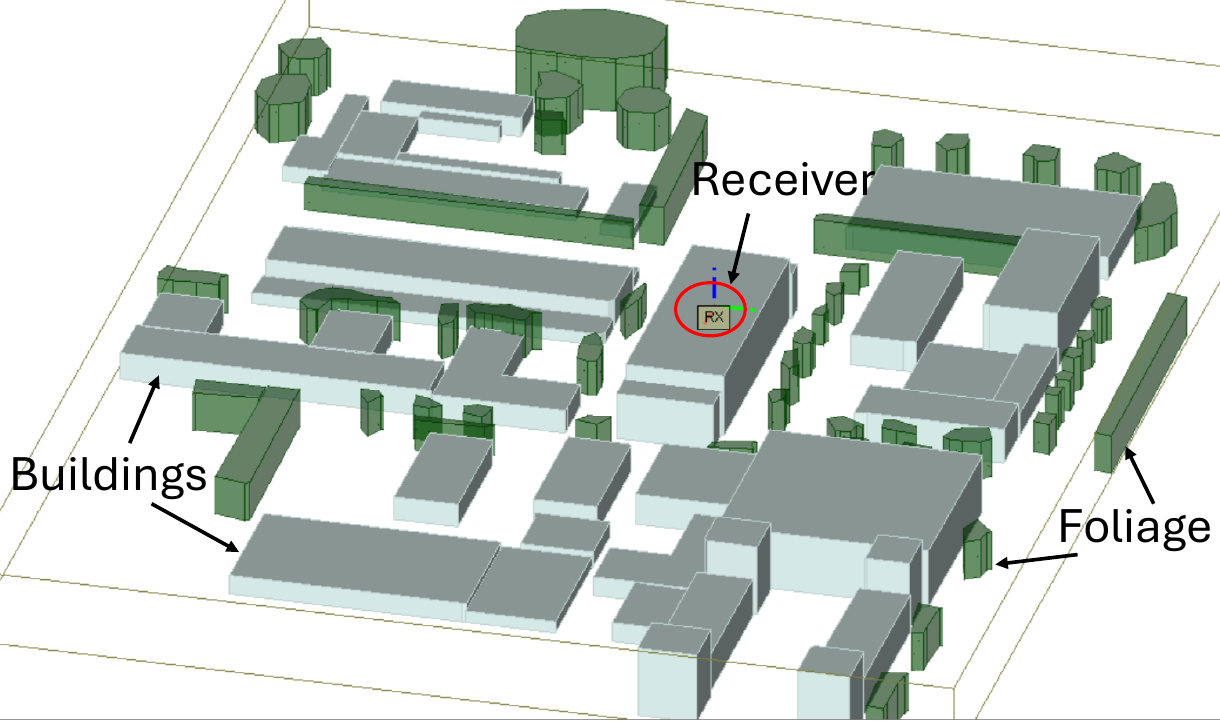} 
        \caption{ }
    \end{subfigure}
    \hfill 
    \begin{subfigure}[b]{0.48\textwidth}
        \centering
        \includegraphics[width=\textwidth]{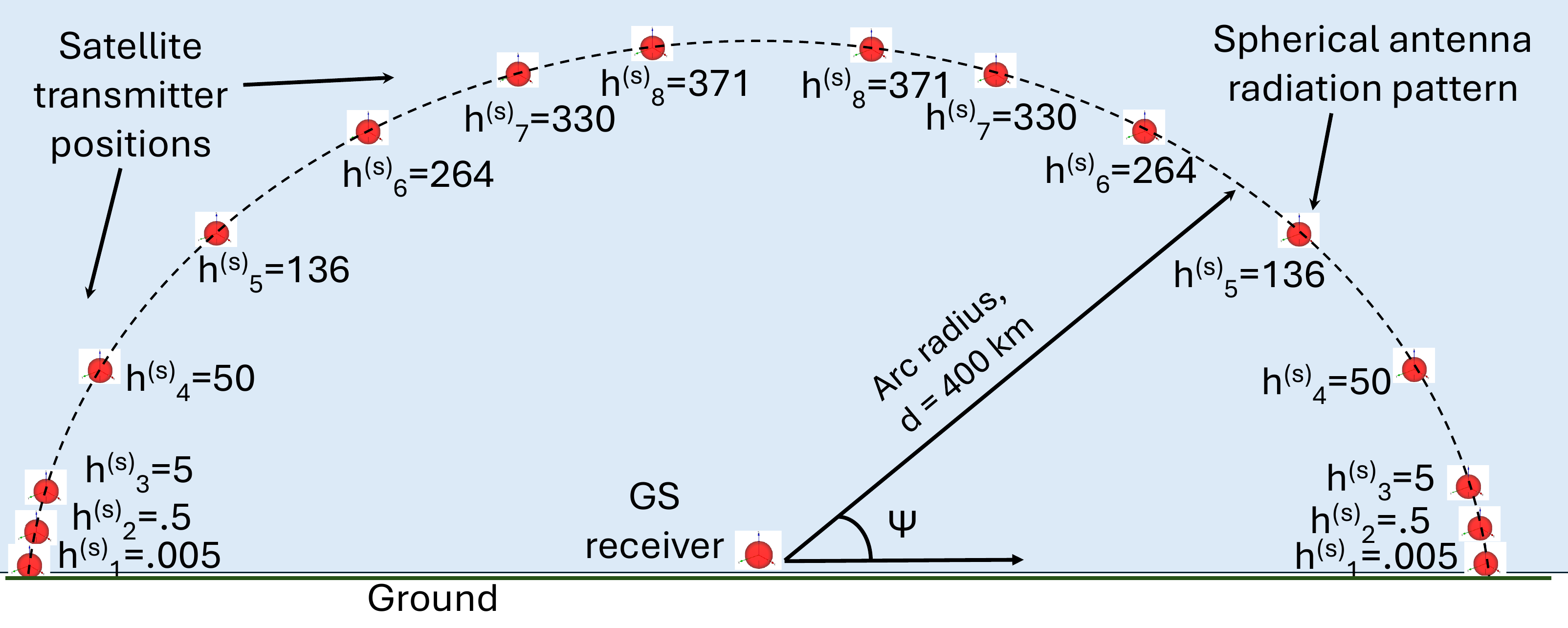}  \caption{}
    \end{subfigure}
    \caption{ (a) GS receiver at the Edison Building and surrounding area, Aarhus University; (b) satellite pass following a circular trajectory directly over the GS with an arc radius of $400$~km, showing multiple satellite elevation positions (in kilometers).}
    \label{Fig:sat_scenario}
\end{figure}

There is significant literature on satellite-to-ground propagation channel models for geostationary and traditional low earth orbit~(LEO). However, there is limited propagation channel literature for small satellites, including CubeSats and microsatellites. Small satellites have shorter visibility windows, smaller antenna gains and lower transmit power, resulting in weak multipath components~(MPCs), and a higher likelihood of antenna misalignment due to limited onboard attitude determination and control capabilities compared to traditional LEO satellites. X-band is increasingly adopted in small satellite missions due to its available bandwidth and moderate atmospheric attenuation. However, existing propagation models offer limited insight into key factors such as elevation-dependent fading dynamics, site-specific scatterers, phased-array beam misalignment, and the effects of clouds, rain, and snow. This work addresses these gaps by providing a detailed characterization of the small satellite propagation channel in the X-band.

In \cite{lit_nasa}, an S-band uplink and X-band downlink CubeSat system was demonstrated, emphasizing dynamic link budgets, antenna testing, and performance optimization. In \cite{lit_fsm}, a finite-state Markov chain model for CubeSat uplink, capturing LEO geometry, Doppler, and urban multipath effects, was proposed and validated via hardware-in-the-loop simulations. Similarly, \cite{fsm_nestor} analyzed queuing delays in LEO satellite-to-ground links using Markov models, showing that short buffers minimized packet loss and delay in S and Ka bands. The Earth-space channel at W-band was characterized in \cite{cubesat_exp}, considering atmospheric effects like attenuation, scintillation, and depolarization. In \cite{rain_leo}, free-space loss and atmospheric attenuation including ionospheric, tropospheric, and local components were analyzed, rain attenuation was identified as a major factor influenced by rainfall rate, location, and path length.

This work presents a novel high-resolution channel model for the satellite-to-ground downlink at X-band~(also at the lower portion of $3$GPP FR3 band), specifically for small LEO satellites (e.g., CubeSats and microsatellites) in suburban campus environments. Elevation-dependent large- and small-scale fading was observed during satellite passes, with rain causing greater attenuation than snow and clouds. Additional differences in attenuation were noted between single-antenna and phased-array configurations. A $400$~km pass exhibited a higher number of MPCs and their clustering, and increased root-mean-square delay spread~(RMS-DS) and angular spread compared to a $500$~km pass. While the methodology is designed to be generalized for various satellite and ground station~(GS) configurations, we validate it using a specific example. The chosen GS is located at the Edison Building, Aarhus University, Denmark, and its surrounding environment is modeled in Wireless InSite ray-tracing software, as shown in Fig.~\ref{Fig:sat_scenario}(a). During the optimal overhead passes, the satellite follows a north-to-south trajectory along a circular arc with a radius of $400$ and $500$~km, centered at the GS. Fig.~\ref{Fig:sat_scenario}(b) shows the scenario with arc radius of $400$~km and different elevation points of the satellite observed from the GS at respective elevation angles $\Psi$.



\section{LEO Small-Scale Fading Model}  \label{Section:Small_scale_fading}
This section presents small-scale fading statistics, RMS delay and angular spreads, and a MPC clustering method for the satellite-to-ground channel.

\subsection{Small-Scale Fading} \label{Section:small_scale}
From ray-tracing channel simulations, the number and characteristics of MPCs vary with satellite elevation and distance from the GS. Significant shadowing near the horizon is observed due to scatterers around the GS. The satellite-to-ground propagation channel at a given satellite elevation angle $\Psi$ shown in Fig.~\ref{Fig:sat_scenario}(b) can be modeled as 
\begin{equation}
\begin{split}
H(t, \Psi) &= \sum_{i=1}^{N_{\Psi}}\alpha_i\big(\Psi\big)\delta\big(t-\tau_i(\Psi)\big) \exp{\big(j\chi_i}\big) \\
&\quad \delta\Big(\pmb{\phi}^{(\rm S)} - \phi_i^{(\rm S)}(\Psi)\Big)\delta\Big(\pmb{\theta}^{(\rm S)} - \theta_i^{(\rm S)}(\Psi)\Big) \\
&\quad \delta\Big(\pmb{\phi}^{(\rm GS)} - \phi_i^{(\rm GS)}(\Psi)\Big)\delta\Big(\pmb{\theta}^{(\rm GS)} - \theta_i^{(\rm GS)}(\Psi)\Big)
\end{split}
\label{Eq:channel_response}
\end{equation}
where $N_{\Psi}$ is the total number of MPCs at elevation angle $\Psi$, $\delta$ is the Kroneker delta function, $\alpha_i$ is the complex amplitude of the $i^{\rm th}$ MPC, $\tau_i$ is delay of the $i^{\rm th}$ MPC, $\chi_i$ is the phase of $i^{\rm th}$ MPC, $\pmb{\phi}^{(\rm S)}$ represents the azimuth vector at the satellite transmitter~(TX), the elevation vector at the satellite is represented by $\pmb{\theta}^{(\rm S)}$, the azimuth and elevation angles at the satellite for $i^{\rm th}$ MPC are represented by $\phi_i^{(\rm S)}$, and $\theta_i^{(\rm S)}$, respectively, the azimuth and elevation vectors at the GS receiver~(RX) are represented by $\pmb{\phi}^{(\rm GS)}$ and $\pmb{\theta}^{(\rm GS)}$, respectively. Finally, the azimuth and elevation angles at the GS for $i^{\rm th}$ MPC is given by $\phi_i^{(\rm GS)}$, and $\theta_i^{(\rm GS)}$, respectively. MPCs arise from ground reflections and both reflections and diffractions off buildings during the satellite pass. 

 Over the satellite pass, the small-scale fading can be modeled as shadowed Rician near ascending and descending horizons, and as Rician and dominant line-of-sight~(LOS) model above the horizon. Let $\Psi_2$ denote the elevation angle that separates the near-horizon and above-horizon regions. The shadowed Rician fading observed near the ascending and descending horizons, for $\Psi < \Psi_2$, is expressed as
\begin{align}
f_R(r) &= \frac{2 r (K+1)}{\Omega} \exp\left(-\frac{K + m}{K+1}\right) 
I_0\left(2 r \sqrt{\frac{m K}{\Omega (K+1)}}\right) \nonumber  \\ &\, {}_1F_1\left(m; 1; -\frac{K + m}{K+1} \frac{r^2}{\Omega}\right),  \label{Eq:Shad_Kfact}
\end{align}
where $f_R(r)$ is the probability density function of the received signal amplitude $r$, $K$ is the Rician $K$-factor given as
\begin{align}
    K = \frac{\big|\alpha_1^{(\rm s)}\big|^2}{\sum_{i=2}^{N_{\Psi}}\big|\alpha_i\big|^2},
\end{align}
where $\alpha_1^{(\rm s)}$ is the shadowed LOS component, $m$ is the Nakagami shape parameter, $\Omega$ is the average received power, $I_0(\cdot)$ is the modified Bessel function of the first kind (order zero), and ${}_1F_1(\cdot)$ is the confluent hypergeometric function of the first kind.

When the satellite is above the horizon corresponding to $\Psi\geq \Psi_2$, and there are MPCs observed in addition to the unshadowed LOS, we can model the small scale fading using the Rician distribution given as
\begin{align}
f_R(r) &= \frac{2 r (K+1)}{\Omega} \exp\left(-K - \frac{(K+1) r^2}{\Omega}\right) \nonumber \\
&\quad \times I_0\left(2 r \sqrt{\frac{K (K+1)}{\Omega}}\right).
\label{eq:rician_pdf_split}
\end{align}
When there is only a LOS component present, then there are no amplitude fluctuations due to small-scale fading, and we have a deterministic single-path model. 

\subsection{RMS-DS and Angular Spreads}
The RMS-DS, denoted as \( \tau_{\mathrm{rms}} \), at a given satellite elevation angle $\Psi$ and distance $d$, can be computed as
\begin{equation}
\tau_{\mathrm{rms}}(\Psi, d) = \sqrt{ \frac{ \sum_{i=1}^{N_{\Psi}} P_i{(\Psi,d)} \bigg(\tau_i (\Psi,d) - \bar{\tau}(\Psi,d)\bigg)^2 }{ \sum_{i=1}^{N} P_i{(\Psi,d)} } }, \label{Eq:RMS_DS}
\end{equation}
where $P_i{(\Psi,d)}$ is the received power of the $i^{\rm th}$ MPC at elevation angle $\Psi$ and distance $d$, and $\bar{\tau}(\Psi,d)$ is the mean excess delay, given by
\begin{equation}
\bar{\tau}(\Psi,d) = \frac{ \sum_{i=1}^{N_{\Psi}} P_i{(\Psi,d)} \tau_i(\Psi,d) }{ \sum_{i=1}^{N_{\Psi}} P_i{(\Psi,d)} }.
\end{equation}

The angular spread of MPCs can be obtained in both the azimuth and elevation planes at the satellite and GS. The azimuth angular spread at the satellite for a given $\Psi$ is denoted by $\sigma^{(\rm S)}_{\phi}{(\Psi)}$ (in degrees) and is given as 
\begin{align}
  & \sigma^{(\rm S)}_{\phi}{(\Psi)} = \frac{180}{\pi} \sqrt{ -2 \ln l{(\Psi)} }, \label{Eq:Az_spread}
 \end{align}
   where the mean resultant length \( l{(\Psi)} \), is given as 
   \begin{align}
     l{(\Psi)} = \frac{1}{N_\Psi} \sqrt{ \left( \sum_{i=1}^{N_\Psi} \cos\phi^{(\rm S)}_i{(\Psi)} \right)^2 + \left( \sum_{i=1}^{N_\Psi} \sin\phi^{(\rm S)}_i{(\Psi)} \right)^2 }.
\end{align}
The elevation angular spread at the satellite, $\sigma^{(\rm S)}_{\theta}{(\Psi)}$, is
\begin{align}
\sigma^{(\rm S)}_{\theta}{(\Psi)} = \sqrt{ \frac{1}{N_\Psi} \sum_{i=1}^{N_\Psi} \bigg( \theta^{(\rm S)}_i{(\Psi)} - \bar{\theta}^{(\mathrm{S})}(\Psi) \bigg)^2 }, \label{Eq:El_spread}
\end{align}
where $\bar{\theta}^{(\mathrm{S})}(\Psi)$ is the mean elevation angle. Similarly, the angular spread of the MPCs at the GS are computed.

\subsection{Clustering of MPCs}  \label{Section:clustering}
In LEO satellite-to-ground channels, MPCs often exhibit clustering behavior due to spatial and temporal correlation among scattered paths~\cite{wahab_predict}. The delay and spatial information of $i^{\rm th}$ MPC at $\Psi^{\rm th}$ elevation angle is represented by a five dimensional feature vector and features are collected into a matrix as follows:
\begin{align}
\mathbf{y}_i(\Psi) = 
\begin{bmatrix}
\tau_i(\Psi) \\ 
\phi^{\text{(S)}}_i(\Psi) \\ 
\theta^{\text{(S)}}_i(\Psi) \\ 
\phi^{\text{(GS)}}_i(\Psi) \\ 
\theta^{\text{(GS)}}_i(\Psi)
\end{bmatrix}^T
\in \mathbb{R}^5, & \mathbf{Y}(\Psi)&= 
\begin{bmatrix}
\mathbf{y}_1(\Psi) \\ 
\mathbf{y}_2(\Psi) \\ 
\vdots \\ 
\mathbf{y}_{N}(\Psi)
\end{bmatrix}
\in \mathbb{R}^{N \times 5}.  \label{Eq:channel_parameters}
\end{align}
Each feature column in $\mathbf{Y}(\Psi)$ is normalized to have zero mean and unit variance:
\begin{align}
 \tilde{\mathbf{Y}}(\Psi) = \text{normalize}(\mathbf{Y}(\Psi)). \label{Eq:normalized_Y}
\end{align}
To account for the periodic nature of azimuth angles, the azimuth components are mapped to the unit circle using $\phi\xrightarrow{}(\sin \phi, \cos \phi)$ before normalization.

Density-based spatial clustering of applications with noise~(DBSCAN)~\cite{dbscan} is applied to the normalized matrix $\tilde{\mathbf{Y}}(\Psi)$ using Euclidean distance, neighborhood radius $\xi$, and minimum points $\zeta$. A point $y_i$ is a core point if the number of neighbors within $\xi$ satisfies 
\begin{align}
|M_\xi(y_i)| \geq \zeta, ~\rm{where}~ M_\xi(y_i) = \{ y_j : \| y_j - y_i \| \leq \xi \}
\end{align}
is the set of all points $y_j$ within a distance of $\xi$ of $y_i$. Clusters are formed from core points by including density-reachable points. Points not reachable from any cluster are labeled as noise and the total number of clusters is $C_{\Psi}$. DBSCAN is well-suited for satellite channel clustering as it automatically detects an unknown number of multipath clusters, and identifies weak MPCs as noise.

\section{LEO Large-Scale Fading Model}  \label{Section:Large_scale_fading}
This section models large-scale fading in LEO satellite links considering total link attenuation from path loss, hardware losses, antenna misalignment, and atmospheric effects. Based on the geometric ray propagation, the large-scale fading observed by the GS at an arc radius $d$ and elevation angle $\Psi$ shown in Fig.~\ref{Fig:sat_scenario}(b) for a given antenna polarization is expressed as 
\begin{align}
  L^{(\mathrm{tot})}(\Psi,d)\ [\mathrm{dB}] &=  
  P^{(\mathrm{TX})} - P^{(\mathrm{RX})}(\Psi,d), \label{eq:PL_total} 
\end{align}
where $L^{(\mathrm{tot})}(\Psi,d)$ is the total link attenuation, $P^{(\rm TX)}$ is the transmit power, $P^{(\rm RX)}\big(\Psi,d\big)$ is the received power given as 
\begin{flalign}
  P^{(\mathrm{RX})}(\Psi,d)\ [\mathrm{dB}] &= P^{(\mathrm{coh})}(\Psi,d) - L^{\rm (hd)} - L^{(\mathrm{am})}(\Delta \phi, \Delta \theta)   \nonumber \\
  &\quad - L^{(\mathrm{atm})}, & \label{eq:RX_power}
\end{flalign}
where $P^{(\rm coh)}\big(\Psi,d\big)$ is the coherent received power from the MPCs including LOS given as 
\begin{align}
  P^{(\mathrm{coh})}(\Psi,d) = 10\log_{10}\bigg(\sum_{i=1}^{N_{\Psi,d}} \Bigl|\alpha_i(\Psi,d)\exp(j\chi_i)\Bigr|^2\bigg). \label{eq:coh_power}
\end{align}
When omnidirectional free space propagation is considered, $P^{(\mathrm{coh})}(d) = 20\log_{10}\big(\frac{\lambda}{4\pi d}\big),$ where
$\lambda$ is the wavelength. Also, from (\ref{Eq:RMS_DS}), $P_i{(\Psi,d)} = \big|\alpha_i(\Psi,d)\exp(j\chi_i)\big|^2$. In (\ref{eq:RX_power}), $L^{\rm (hd)}$ is the attenuation due to hardware including cable and antenna feed losses, $L^{(\rm am)}(\Delta \phi, \Delta \theta)$ is the attenuation in dB due to antenna beam misalignment by $\Delta \phi$, and $\Delta \theta$ in the azimuth and elevation planes, respectively, at the GS, $L^{(\rm atm)}$ is the attenuation due to adverse weather conditions and atmospheric attenuation including gaseous attenuation and tropospheric scintillation at X-band for a given polarization~\cite{gas_atten,trop_scin}.

The attenuation due to antenna beam misalignment compared to when the antennas are aligned at the boresight at the satellite and the GS is given as 
\begin{align}
  L^{(\mathrm{am})}(\Delta \phi, \Delta \theta)\ [\mathrm{dB}] &= G^{(\mathrm{GS})}(\phi, \theta) \nonumber \\
  &\quad - G^{(\mathrm{GS})}(\phi+\Delta \phi, \theta+\Delta \theta), & \label{Eq:PL_misalign}
\end{align}
where $G^{(\rm GS)}\big(\phi, \theta \big)$ is the gain of the antenna at the GS when aligned with the satellite antenna boresight, and $G^{(\rm GS)}\big(\phi+\Delta \phi, \theta+\Delta \theta \big)$ is the gain of the antenna at GS when misaligned by $\Delta \phi$ and $\Delta \theta$ from the boresight of the satellite antenna.  

The attenuation in adverse weather conditions, e.g., dense clouds, rain, and snow increases the overall link attenuation in (\ref{eq:PL_total}). The attenuation in [dB] due to rain, clouds, snow, and atmosphere is given as 
\begin{align}
 L^{(\mathrm{atm})} = L^{(\mathrm{rn})}(\Psi) + L^{(\mathrm{cl})}(\Psi) + L^{(\mathrm{sn})}(\Psi) + L^{\rm (fx)},  \label{Eq:PL_weather}
\end{align}
where $L^{(\rm rn)}\big(\Psi\big)$, $L^{(\rm cl)}\big(\Psi\big)$, and $L^{(\rm sn)}\big(\Psi\big)$ are the attenuation due to rain, dense clouds, and snow, respectively, and are dependent on the elevation angle $\Psi$ for a given antenna polarization, and $L^{\rm (fx)}$ is the fixed atmospheric attenuation. 

The attenuation due to rain for circular polarization $ L^{(\rm rn)}\big(\Psi\big)$ is given as~\cite{ITU_rain} 
\begin{align}
  L^{(\mathrm{rn})}(\Psi) = \gamma_{\mathrm{R}} \cdot L_{\mathrm{E}} + \beta, & \quad L_{\mathrm{E}}(\Psi) = L_{\mathrm{G}} \frac{r_{0.01}}{\cos(\Psi)}, & \label{Eq:rain_att}
\end{align}
where $\beta$ is the polarization-dependent attenuation constant, $r_{0.01}$ is the horizontal reduction factor for a rainfall rate of $0.01\%$ exceedance~(25 mm/h) given as 
\begin{align}
  r_{0.01} = \frac{1}{1 + 0.78 \sqrt{\frac{L_G \gamma_{\mathrm{R}}}{f_{\mathrm{c}}}} - 0.38 \left( 1 - e^{-2 L_{\mathrm{G}}} \right)},  \label{Eq:r_rain}
\end{align}
where $f_{\rm c}$ is the center frequency, the $ \gamma_{\rm R}$ and $L_{\rm G}$ can be obtained as follows:
\begin{align}
  \gamma_{\mathrm{R}} = k^{(\mathrm{rn})} R_{0.01}^\epsilon, & \quad L_{\mathrm{G}} = L_{\mathrm{s}} \cos(\Psi), & \label{Eq:gamma_LG}
\end{align}
where $k^{(\rm rn)}$ and $\epsilon$ are rain attenuation coefficients, $R_{0.01}$ is the rainfall rate for $0.01\%$ exceedance, and 
\begin{align}
 L_{\mathrm{s}} = \sqrt{\frac{2(h^{(\mathrm{R})} - h^{(\mathrm{GS})}) R_{\mathrm{e}}}{\sin^2(\Psi) + \frac{2(h^{(\mathrm{R})} - h^{(\mathrm{GS})})}{R_{\mathrm{e}}}}} + \frac{h^{(\mathrm{R})} - h^{(\mathrm{GS})}}{\sin(\Psi)},  \label{Eq:slant_range}
\end{align}
where $h^{(\rm R)}$ is the rain height, $h^{(\rm GS)}$ is the GS height, and $R_{\rm e}$ is the effective earth radius. Similarly, the attenuation during rain for other antenna polarization can be obtained.

The attenuation due to propagation through the clouds for circular polarization is given as~\cite{ITU_clouds}
\begin{align}
 L^{(\rm cl)}\big(\Psi\big) = k^{(\rm cl)}TM\frac{1}{\sin(\Psi)},  \label{Eq:cloud_atten}
\end{align}
where $k^{(\rm cl)}$ is the specific attenuation constant for clouds and will be different for vertical and circular polarization, $T$ is clouds thickness in km, and $M$ is the liquid water density. Similar to rain and cloud attenuation, the attenuation due to snow can be represented as follows:
\begin{align}
 L^{(\rm sn)}\big(\Psi\big) = k^{(\rm sn)}S\frac{h^{(\rm S)}}{\sin(\Psi)},  \label{Eq:atten_snow}
\end{align}
where $k^{(\rm sn)}$ is specific attenuation coefficient for snow (dB/km per mm/h), S is the snowfall rate in mm/h, and $h^{(\rm S)}$ is the effective snow height in km.  

\section{Simulation Setup}    \label{Section:Simulations}
This section describes the Wireless InSite simulation setup, compares results with $3$GPP NTN channel profiles, and provides antenna gain spatial filtering of ray-traced MPCs.
\subsection{Ray-Tracing Software and $3$GPP Channel Profile}
The simulations are carried out using Wireless InSite ray-tracing software. The GS is located at the Edison Building, Aarhus University, Denmark. The building over which the GS is mounted and surrounding structures and foliage are created using the Wireless InSite software shown in Fig.~\ref{Fig:sat_scenario}(a). The ray-tracing simulation parameters are provided in Table~\ref{Table:sim_par}. The values of the simulation parameters for clouds, rain, and snow attenuation, discussed in Section~\ref{Section:Large_scale_fading} are also provided in Table~\ref{Table:sim_par}, are according to local conditions in Northern Europe. 

$3$GPP NTN channel model~\cite{3gpp} is also used for comparison with the ray-tracing simulations. Different tapped delay line~(TDL) channel profiles, $f^{(\rm ch)}$, are used at different satellite elevations given as 
\begin{align}
 f^{(\rm ch)}(\Psi) =
\begin{cases}
\text{NTN-TDL-A}, & \Psi < \Psi_1 \\
\text{NTN-TDL-B}, & \Psi_1 \leq \Psi < \Psi_2 \\
\text{NTN-TDL-C}, & \Psi_2 \leq \Psi,
\end{cases}~.  \label{Eq:3GPP}
\end{align}
Compared to two angular regions, ($\Psi<\Psi_2$ and $\Psi \geq \Psi_2$) for small scale fading statistics in Section~\ref{Section:small_scale}, three angular regions are defined for the $3$GPP NTN satellite to ground propagation channel. All three channel profiles contain non-LOS~(NLOS) components. The NTN-TDL-A profile is applied for elevation angles below $\Psi_1$ where the LOS path is blocked by surrounding scatterers. Between elevation angles $\Psi_1$ and $\Psi_2$ the NTN-TDL-B profile is used to represent a shadowed LOS condition. For elevation angles above $\Psi_2$, a clear and strong LOS component is observed and modeled using the NTN-TDL-C profile. Based on typical values observed in satellite propagation studies, $\Psi_1=10^{\circ}$ and $\Psi_2=15^{\circ}$, and the standard deviation of the log-normal shadowing component is set to $8$~dB, $6$~dB, and $4$~dB for the NTN-TDL-A, NTN-TDL-B, and NTN-TDL-C profiles, respectively, indicating decreasing variability with increasing satellite elevation angle. Additionally, antenna gains are added as deterministic offsets to the stochastic $3$GPP channel model.

\begin{table}[!t]
	\begin{center}
		\caption{Parameters for simulations.}\label{Table:sim_par}
		\resizebox{\columnwidth}{!}{
        \begin{tabular}{@{} |P{7.2cm}|P{3.0cm}| @{}}
			\hline
			\textbf{Parameter}&\textbf{Parameter value}\\			
			\hline
		    Center frequency, $f_{\rm c}$& $10$~GHz \\
            \hline
            Antenna radiation pattern & Spherical \\
            \hline
            Antenna polarization& Left hand circular \\
            \hline
            Transmit power, $P^{(\rm TX)}$
            & $30$~dBm \\
            \hline
            Permittivity of ground
            &  $3.5$\\
            \hline
            Permittivity of scatterer structure
            & $5.31$ \\
            \hline
            Height of GS (RX), $h^{(\rm GS)}$
            & $23$~m \\
            \hline
            Maximum height of the satellite above GS, $h_{\rm max}^{(\rm s)}$
            & $\{400,500\}-.023$ km \\
            \hline
            Foliage
            & Dense deciduous tree \\
            \hline
            Polarization-dependent attenuation constant, $\beta$
            & 3 dB \\
            \hline 
             Rain height, $h^{(\rm R)}=h^{(\rm S)}$
            & $5$ km \\
            \hline 
            Effective radius of the earth, $R_{\rm e}$
            & $6371$ km \\
            \hline
            Rainfall and snow rate, \{$R_{.01}$,~ $S$\}
            & \{32,~4\} mm/h \\
            \hline
            Rain attenuation coefficient, $k^{(\rm rn)}$
            & $0.0363$ \\
            \hline
            Rain attenuation coefficient, $\epsilon$
            & $1.095$\\
            \hline
            Specific attenuation constant for clouds and snow, $\big\{k^{(\rm cl)},~ k^{(\rm sn)}\big\}$
            & \{$0.072$,~ $0.004$\} \\
            \hline
            Clouds thickness, $T$
            & $1.5$~km \\
            \hline
            Liquid water density, $M$
            & $0.35$ g/m$^3$ \\
            \hline
             Hardware and fixed atmospheric attenuation,~\big\{$L^{\rm (hd)}$, $L^{\rm (fx)}$\big\}
            & \{$1.5$, $1.5$\}~dB \\
            \hline
        \end{tabular}
            }
		\end{center}
\end{table}

\begin{figure}[t!]
    \centering
    \begin{subfigure}[b]{0.4925\columnwidth} 
        \centering
        \includegraphics[width=\textwidth]{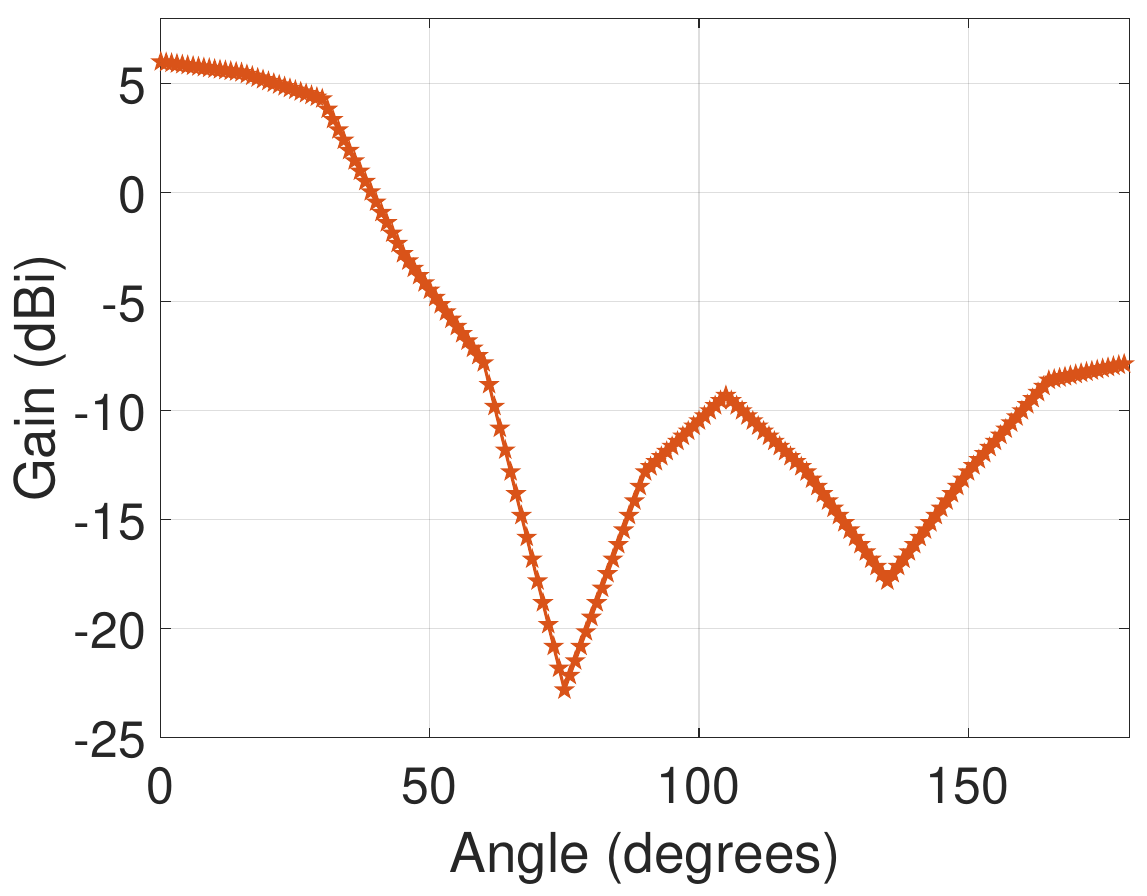}
        \caption{}
    \end{subfigure}
    \hfill 
    \begin{subfigure}[b]{0.4925\columnwidth}
        \centering
        \includegraphics[width=\textwidth]{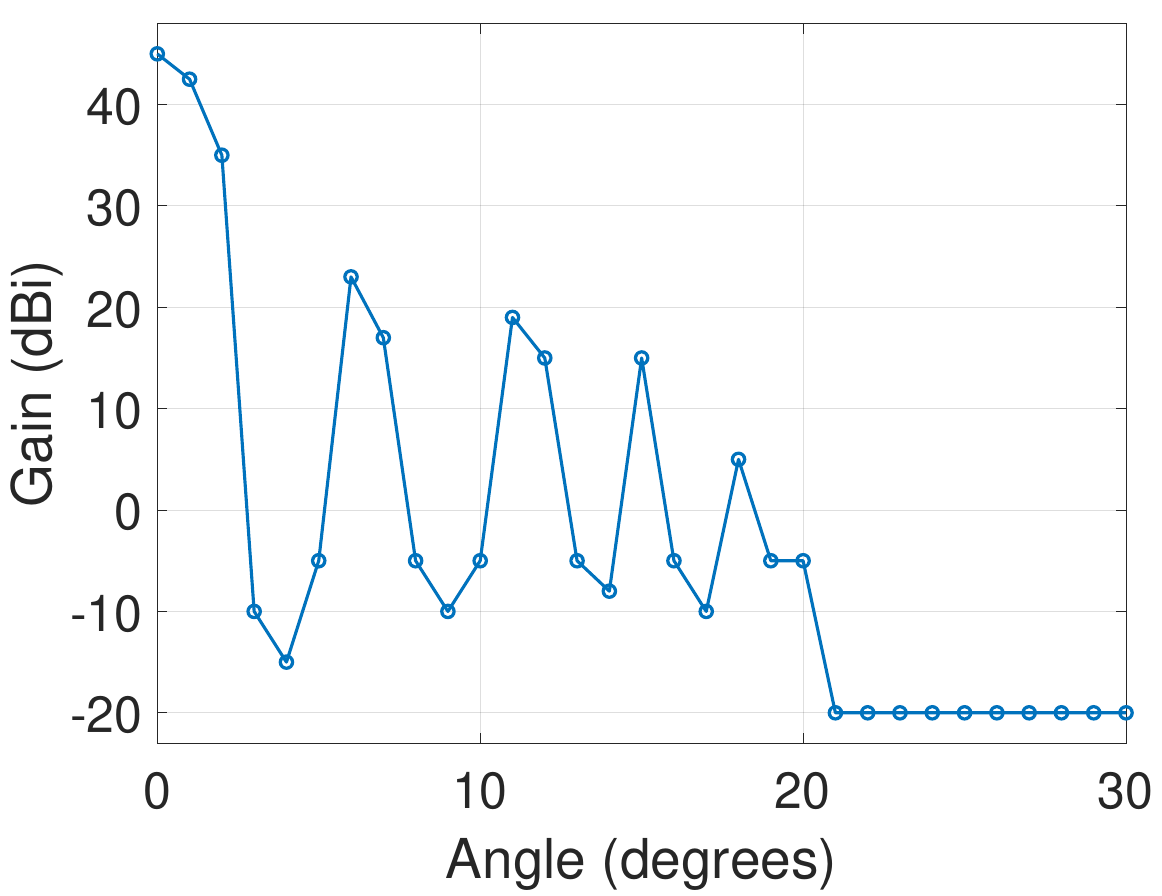}
        \caption{}
    \end{subfigure}
    \begin{subfigure}[b]{0.4925\columnwidth} 
        \centering
        \includegraphics[width=\textwidth]{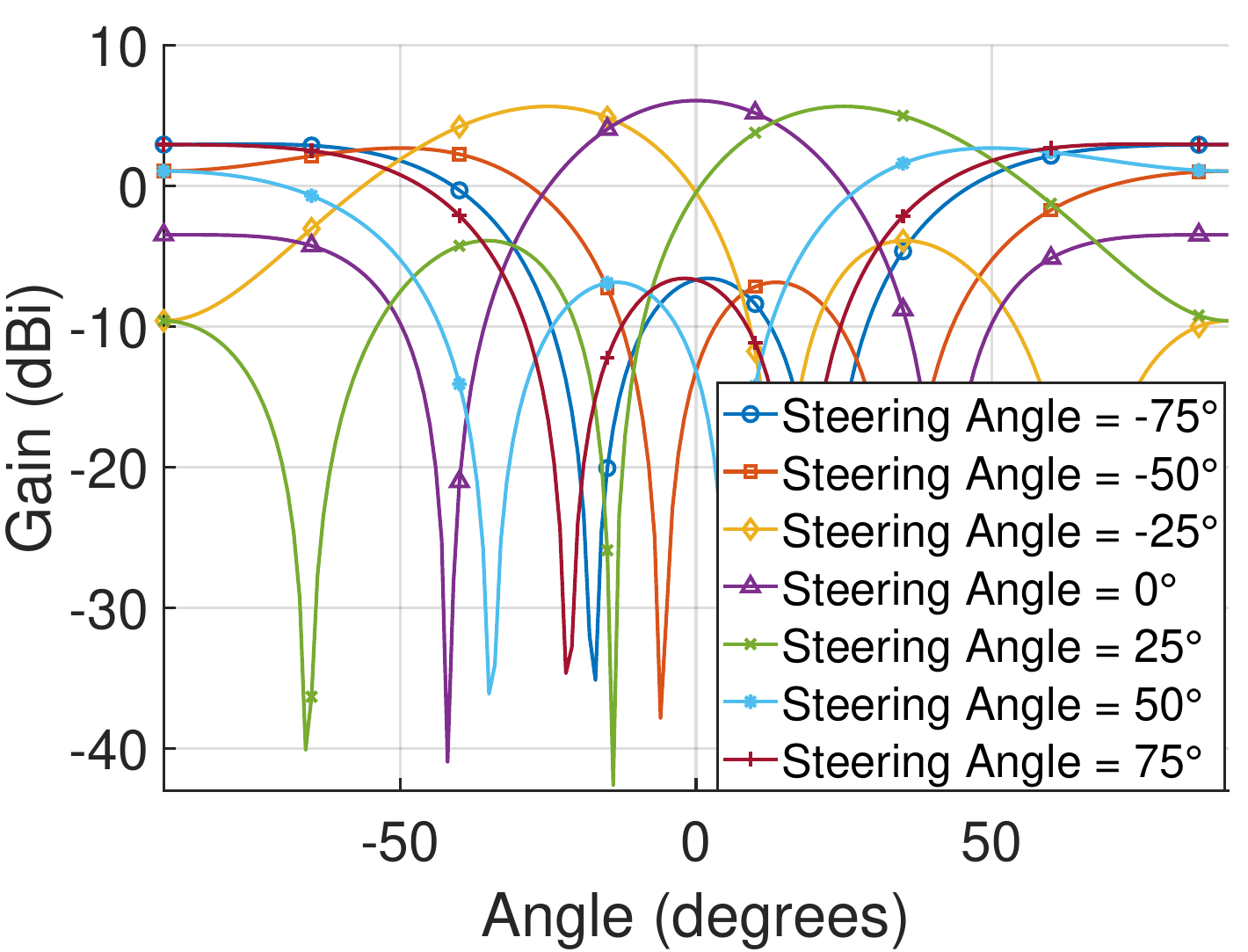}
        \caption{}
    \end{subfigure}
    \hfill 
    \begin{subfigure}[b]{0.4925\columnwidth} 
        \centering
        \includegraphics[width=\textwidth]{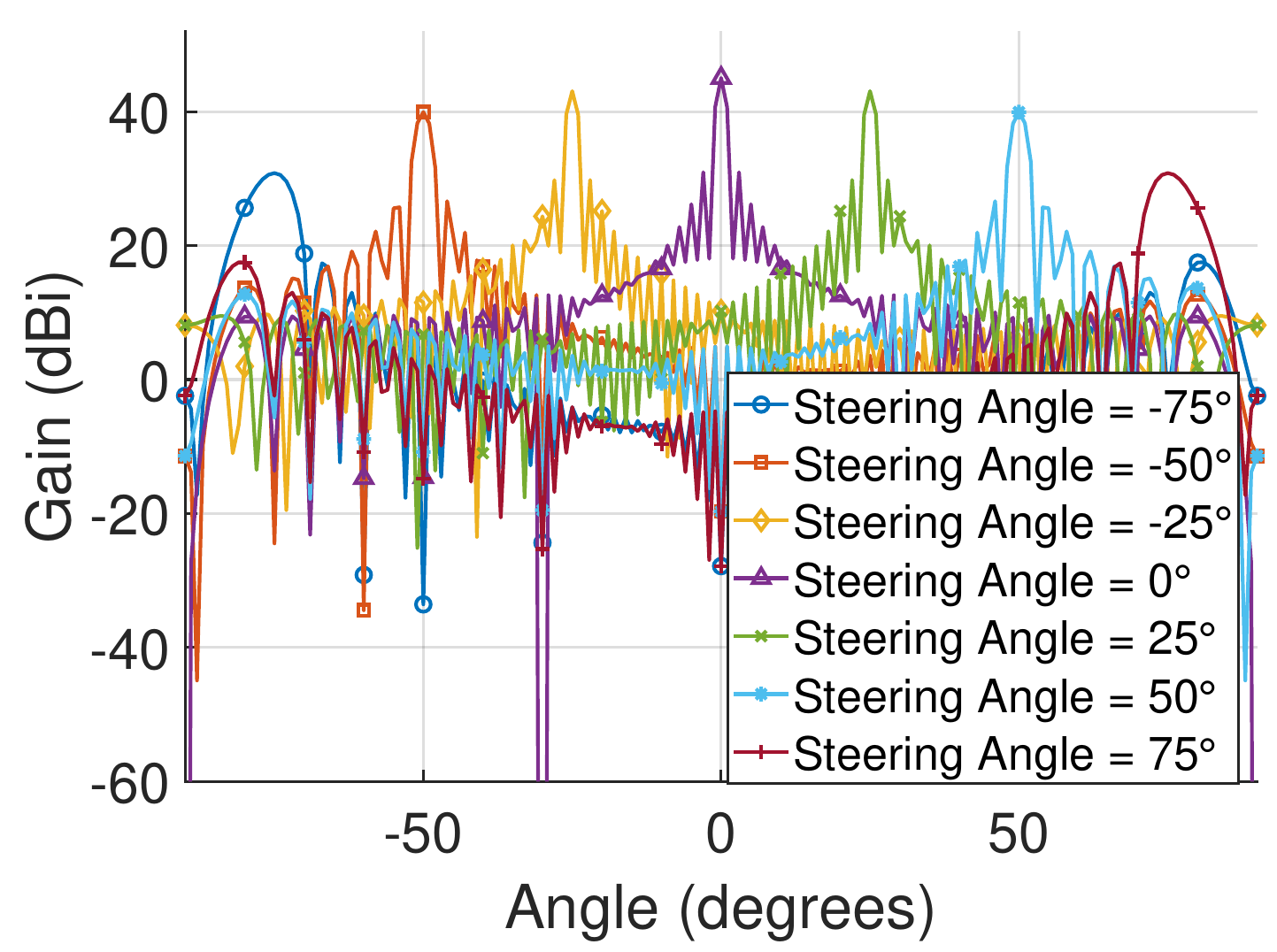}
        \caption{}
    \end{subfigure}
    \caption{Antenna radiation patterns (similar in the azimuth and elevation planes) for: (a) satellite single-antenna; (b) GS single-antenna; (c) satellite phased array at multiple steering angles with $3\times3$ elements; and (d) GS phased array at multiple steering angles with $60\times60$ elements.}
    \label{fig:ant_patt}
\end{figure}

\subsection{Antenna Gain Spatial Filtering}
We have used a spherical antenna radiation pattern for the simulations to optimally capture rays from different spatial positions, as shown in Fig.~\ref{Fig:sat_scenario}(b). We have performed antenna gain spatial filtering~\cite{spatial_filter} on the received rays to map the spherical antenna radiation pattern to the actual antenna radiation pattern at the satellite and GS shown in Fig.~\ref{fig:ant_patt}. In the antenna gain spatial filtering, the antenna gain of each ray is modified from a fixed value (due to spherical radiation pattern) to an antenna gain dependent on its azimuth and elevation angles at the satellite and GS. To model the antenna beam misalignment, each ray is assigned an antenna gain based on the misalignment in the azimuth and elevation planes compared to the boresight alignment~cf.~(\ref{Eq:PL_misalign}). 

\begin{figure}[!t]
	\centering
	\includegraphics[width=0.96\columnwidth]{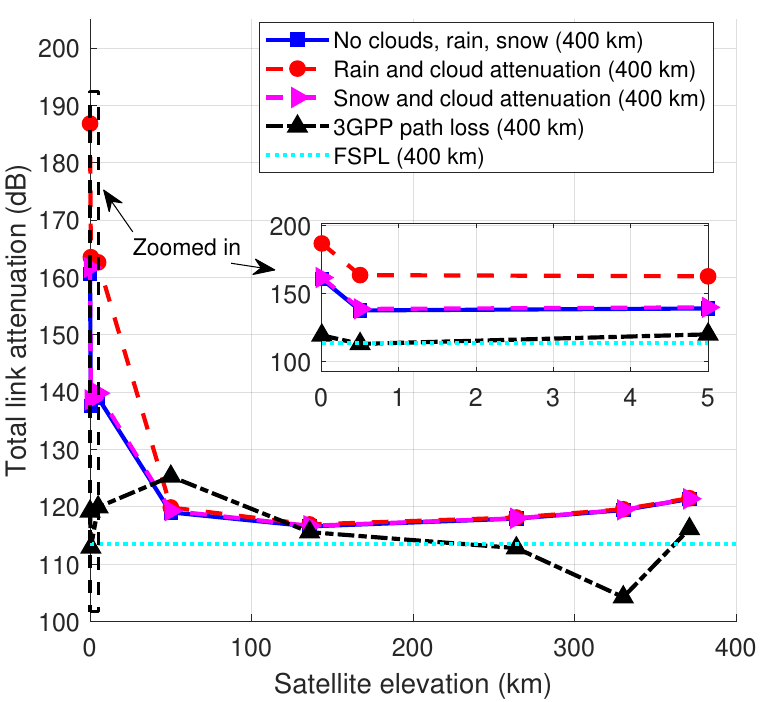}
	\captionsetup{justification=justified, singlelinecheck=false}
	\caption{Total link attenuation for satellite pass at $400$~km under varying weather conditions. Link attenuation from the $3$GPP NTN channel model (\ref{Eq:3GPP}) and FSPL are also provided.}\label{Fig:PL_400_circ}
\end{figure}

\begin{figure}[!t]
	\centering
	\includegraphics[width=0.98\columnwidth]{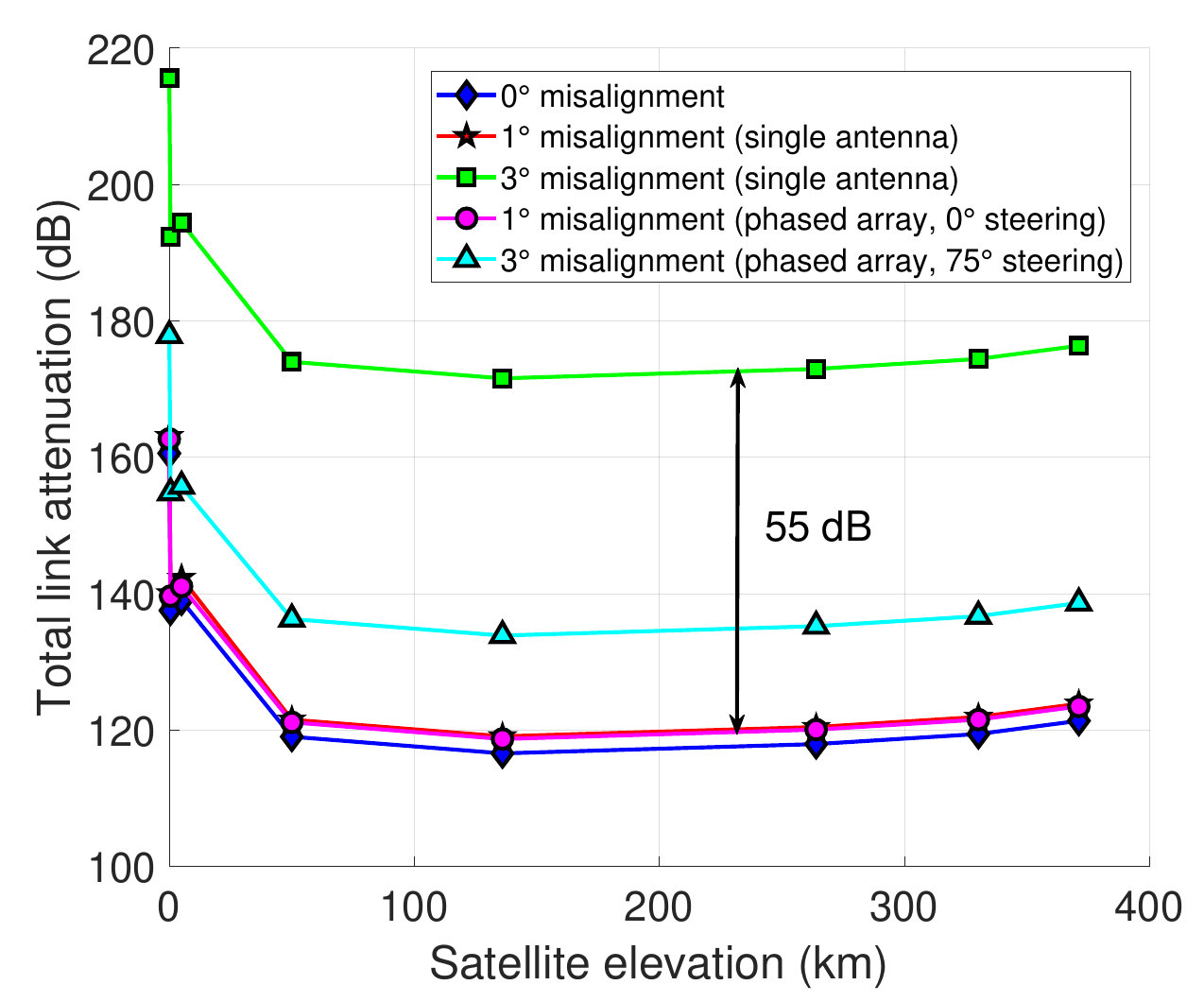}
	\captionsetup{justification=justified, singlelinecheck=false}
	\caption{Total link attenuation for $0^{\circ}$, $1^{\circ}$, and $3^{\circ}$ misalignment in the azimuth and elevation planes for $400$~km pass (without clouds, rain and snow) for single-antenna and phased array. }\label{Fig:PL_misalign}
\end{figure}

\section{Simulation Results}      \label{Section:Results}
In this section, the LEO satellite to ground propagation channel simulation results are presented and discussed. 

\subsection{Link Attenuation}

The satellite elevation angle-dependent link attenuation at a $400$~km pass under various weather conditions (modeled in Section~\ref{Section:Large_scale_fading}) is shown in Fig.~\ref{Fig:PL_400_circ}. Greater attenuation occurs at lower elevations due to LOS shadowing and weather effects~(\ref{Eq:rain_att}, \ref{Eq:cloud_atten}, \ref{Eq:atten_snow}), which diminish at higher elevations. Among weather factors, rain causes the highest attenuation compared to snow and clouds. For comparison, Fig.~\ref{Fig:PL_400_circ} also includes $3$GPP NTN model attenuation~(\ref{Eq:3GPP}), excluding atmospheric, weather, and hardware losses. Above $50$~km elevation, ray-tracing results align within $2–7$~dB of both the $3$GPP model and free-space path loss~(FSPL), with differences due to atmospheric, weather attenuation, hardware loss~(\ref{eq:RX_power}, \ref{Eq:PL_weather}), and MPCs interference. Similar observations were made for the $500$~km satellite pass.

The link attenuation due to antenna beam misalignment modeled in (\ref{Eq:PL_misalign}) for no clouds, rain, and snow conditions at $d=400$~km is shown in Fig.~\ref{Fig:PL_misalign}. The link attenuation for the single-antenna increases proportionally with the misalignment, whereas, for the phased array it is dependent on the steering angle of the beam and corresponding radiation pattern (see Fig.~\ref{fig:ant_patt}). For example, in Fig.~\ref{Fig:PL_misalign}, the link attenuation at $3^\circ$ misalignment at the GS is larger for the single-antenna compared to link attenuation at the same misalignment angle for the phased array due to broader phased array radiation beam at $75^\circ$ steering angle. 

\begin{table}[ht]
\centering
\caption{Number of MPCs and clusters, and $K$-factor at different satellite elevations without rain, snow, or clouds, and with aligned antenna beams for $400$~km and $500$~km passes.}
\label{tab:channel_params}
\renewcommand{\arraystretch}{1.1}
\footnotesize
\setlength{\tabcolsep}{2pt}
\begin{tabular}{|>{\centering\arraybackslash}p{8mm}|
                 >{\centering\arraybackslash}p{8.5mm}|
                 >{\centering\arraybackslash}p{10.2mm}|
                 >{\centering\arraybackslash}p{8mm}||
                 >{\centering\arraybackslash}p{8mm}|
                 >{\centering\arraybackslash}p{8.5mm}|
                 >{\centering\arraybackslash}p{10.2mm}|
                 >{\centering\arraybackslash}p{8mm}|}
\hline
\multicolumn{4}{|c||}{\bfseries $\mathbf{400}$~km Satellite Pass} & \multicolumn{4}{c|}{\bfseries $\mathbf{500}$~km Satellite Pass} \\
\hline
Elev. (km) & \# of MPCs & $K$-factor & \# of clust. & Elev. (km) & \# of MPCs & $K$-factor & \# of clust. \\
\hline
$0.005$  & $2$ & $0.007$   & $0$ & $0.005$    & $2$ & 0.005    & $0$ \\
\hline
$0.5$    & $3$ & $0.11$   & $1$ & $0.5$      & $2$ & 0.087    & $0$ \\
\hline
$5$      & $2$ & $2.26$   & $0$ & $10$       & $2$ & 0.77    & $0$ \\
\hline
$50$     & $2$ & $52.94$  & $0$ & $75$       & $1$ & --       & $0$ \\
\hline
$136$    & $2$ & $210.36$ & $0$ & $150$      & $1$ & --       & $0$ \\
\hline
$264$    & $4$ & $53.50$  & $1$ & $250$      & $1$ & --       & $0$ \\
\hline
$330$    & $3$ & $168.73$ & $1$ & $330$      & $2$ & $158.53$  & $0$ \\
\hline
$371$    & $3$ & $82.21$  & $1$ & $400$      & $2$ & $142.79$  & $0$ \\
\hline
--     & -- & --     & -- & $480$      & $2$ & $197.02$  & $0$ \\
\hline
--     & -- & --     & -- & $500$      & $9$ & $119.93$  & $2$ \\
\hline
\end{tabular}
\end{table}

\subsection{Number of MPCs, Rician $K$-factor, and Clustering}
The number of significant MPCs obtained using ray-tracing simulations observed at various satellite elevations for the $d=400$~km and $d=500$~km passes are provided in Table~\ref{tab:channel_params}. The number of MPCs range from $1$ to $9$ and depend on the satellite's elevation and the surrounding terrain near the GS. A smaller number of MPCs are observed at different satellite elevation points for $500$~km satellite pass compared to $400$~km mainly due to larger attenuation with distance and less resolvable angles of MPCs from scatterers. 


\begin{figure*}[!t]
    \centering
    \begin{subfigure}[t]{0.323\textwidth}
        \centering
        \includegraphics[width=\linewidth]{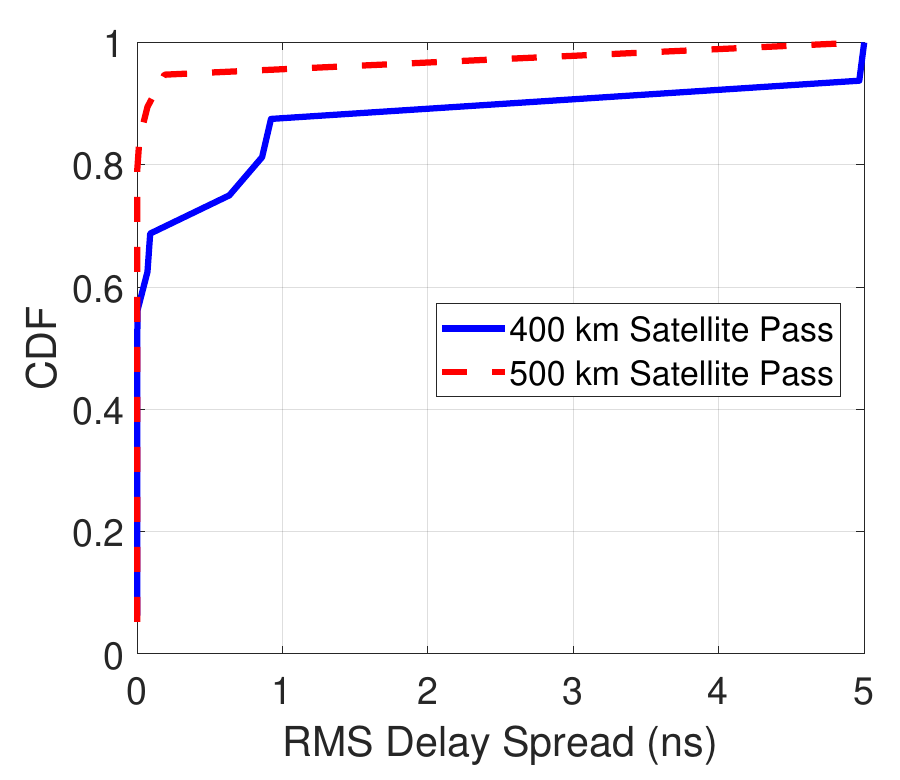}
        \caption{}
        \label{Fig:RMS_DS_CDF}
    \end{subfigure}
    \hfill
    \begin{subfigure}[t]{0.32\textwidth}
        \centering
        \includegraphics[width=\linewidth]{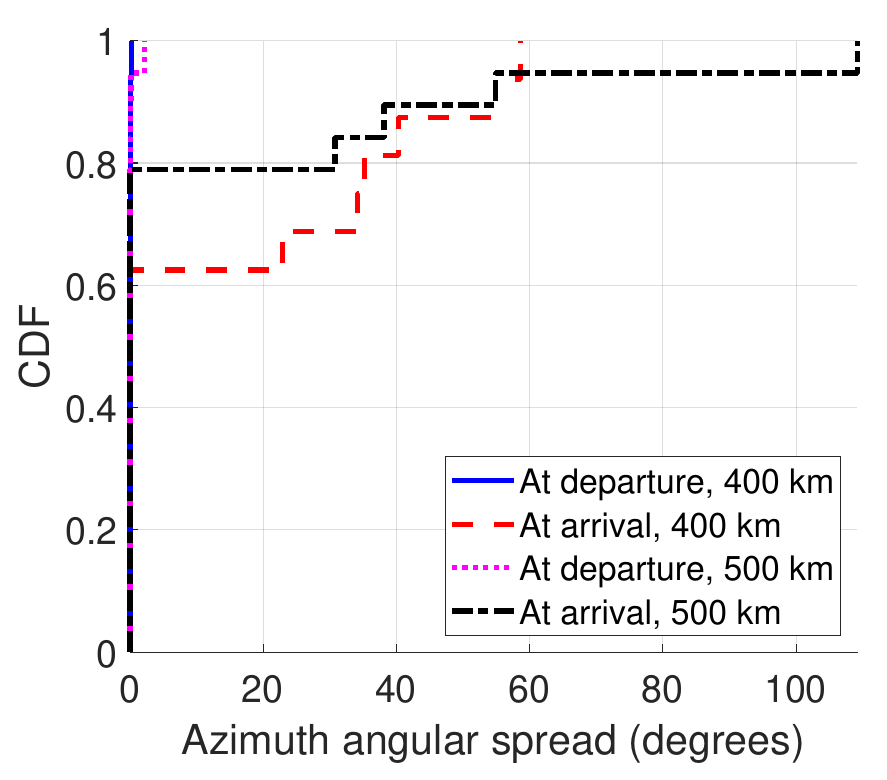}
       \caption{}
        \label{Fig:Az_spread_CDF}
    \end{subfigure}
    \hfill
    \begin{subfigure}[t]{0.32\textwidth}
        \centering
        \includegraphics[width=\linewidth]{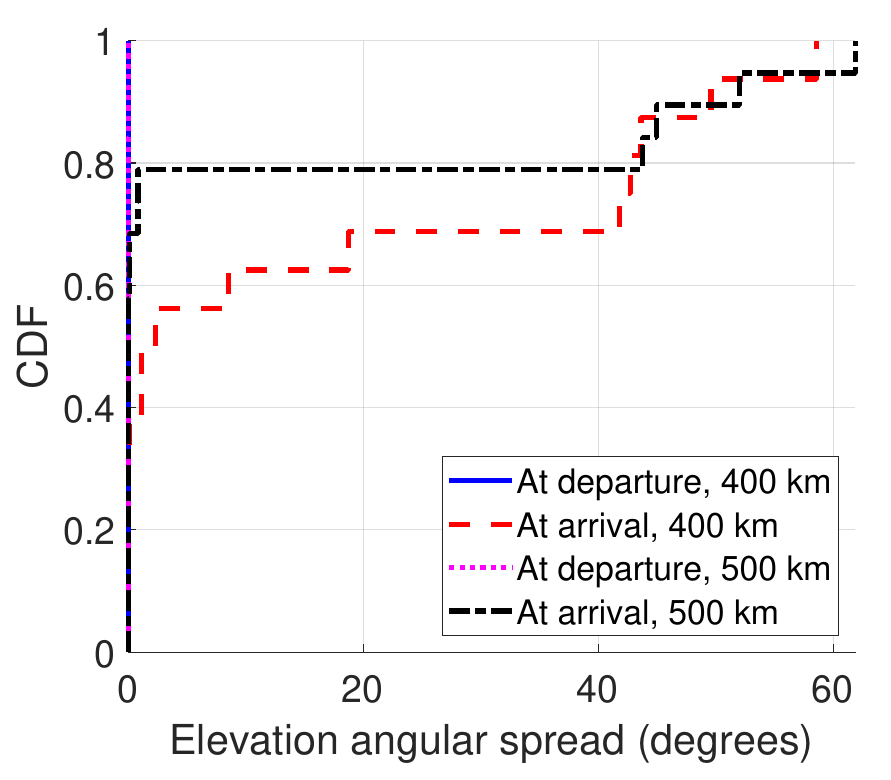}
       \caption{}
        \label{Fig:El_spread_CDF}
    \end{subfigure}
    \captionsetup{justification=justified, singlelinecheck=false}
    \caption{Delay and angular spread statistics of MPCs at different satellite elevations for $400$~km and $500$~km passes.}
    \label{fig:combined_spreads}
\end{figure*}    

The Rician $K$-factor for $400$~km and $500$~km passes is provided in Table~\ref{tab:channel_params}. Lower satellite elevations yield smaller $K$-factors due to LOS shadowing (\ref{Eq:Shad_Kfact}). Beyond certain elevations \big($\Psi_2=\arcsin(\frac{100}{400})$\big), the LOS component is no longer shadowed and its power remains approximately constant, and $K$-factor will depend on the power of NLOS components. Therefore, higher $K$-factor values occur when power of NLOS components is low. At elevations with only a LOS component, the $K$-factor is undefined, and the fading follows a deterministic single-path model~(Section~\ref{Section:small_scale}).


The number of clusters of MPCs~(from Section~\ref{Section:clustering}) for $400$~km and $500$~km satellite passes is provided in Table~\ref{tab:channel_params} for $\zeta=2$ and $\xi=0.3$. Overall, we observe limited number of clusters due to limited number of MPCs. The number of clusters at different satellite elevation angles is higher for the $400$~km pass compared to $500$~km, primarily due to the greater number of MPCs and their stronger correlation in the temporal and angular domains.


\subsection{RMS-DS and Angular Spreads}
Fig.~\ref{fig:combined_spreads}(a) shows the CDF of RMS-DS (\ref{Eq:RMS_DS}) for $400$~km and $500$~km satellite passes. Both scenarios exhibit much lower RMS-DS than typical terrestrial channels, indicating sparse MPCs in satellite-to-ground links. The $500$~km case consistently shows smaller RMS-DS, with over 90\% of values below $1$~ns, compared to $1.2$~ns at $400$~km, due to reduced MPC dispersion at higher altitudes.

Figs.~\ref{fig:combined_spreads}(b) and Figs.~\ref{fig:combined_spreads}(c) show the cumulative distribution functions~(CDFs) of azimuth and elevation angular spreads~(\ref{Eq:Az_spread}, \ref{Eq:El_spread}) at the satellite (departure) and GS (arrival). Angular spreads are significantly smaller at the satellite, indicating directional view. The $500$~km pass results in narrower spreads than $400$~km, due to longer link distances and limited angular visibility.



\section{Conclusions and Future Work}   \label{Section:Conclusions}
This work modeled the satellite-to-ground channel for small satellites in a suburban area using ray-tracing. Significant link attenuation was observed at low satellite elevations due to shadowing and rain. Antenna radiation patterns and misalignment at the GS also impacted losses in both single and phased array configurations. Large- and small-scale fading varied with elevation and distance, with small-scale fading modeled using single path, shadowed Rician, and Rician distributions. Larger number of MPCs and clusters, and greater RMS-DS and angular spreads were observed at $400$~km than at $500$~km. Future work will explore adaptive MODCOD selection based on elevation-dependent link attenuation and channel statistics.


\ifCLASSOPTIONcaptionsoff
  \newpage
\fi

\begin{balance}
\bibliographystyle{IEEEtran}
\bibliography{References}
\end{balance}

\end{document}